\documentclass[12pt,a4paper]{article}
\pdfoutput=1

\usepackage{lineno}  
\usepackage{graphicx}  
\usepackage{xspace} 
\usepackage{color}
\usepackage{colortbl}
\usepackage{amsmath} 
\usepackage{ifthen} 

\newboolean{pdflatex}
\setboolean{pdflatex}{true} 
\newboolean{articletitles}
\setboolean{articletitles}{true} 

\newboolean{uprightparticles}
\setboolean{uprightparticles}{false} 

\usepackage{amssymb}
\usepackage{amsfonts}
\usepackage{upgreek} 

\usepackage{hyperref}    
\usepackage[all]{hypcap} 

\def\ux85 {\mbox{UX85}\xspace}

\ifthenelse{\boolean{uprightparticles}}%
{

 \def\Pmu         {\ensuremath{\upmu}\xspace}

 \def\Ppi         {\ensuremath{\uppi}\xspace}

 \def\Ppsi        {\ensuremath{\uppsi}\xspace}

 \def\PDelta      {\ensuremath{\Delta}\xspace}                 
 \def\PXi      {\ensuremath{\Xi}\xspace}                 
 \def\PLambda      {\ensuremath{\Lambda}\xspace}                 
 \def\PSigma      {\ensuremath{\Sigma}\xspace}                 
 \def\POmega      {\ensuremath{\Omega}\xspace}                 
 \def\PUpsilon      {\ensuremath{\Upsilon}\xspace}                 
 
 \def\PB      {\ensuremath{\mathrm{B}}\xspace}                 
                  
 \def\PD      {\ensuremath{\mathrm{D}}\xspace}

 \def\PJ      {\ensuremath{\mathrm{J}}\xspace}                 
 \def\PK      {\ensuremath{\mathrm{K}}\xspace}

 \def\Pb      {\ensuremath{\mathrm{b}}\xspace}                 
 \def\Pc      {\ensuremath{\mathrm{c}}\xspace}

 \def\Pi      {\ensuremath{\mathrm{i}}\xspace}

 \def\Ps      {\ensuremath{\mathrm{s}}\xspace}

}
{

 \def\Pmu         {\ensuremath{\mu}\xspace}

 \def\Ppi         {\ensuremath{\pi}\xspace}

 \def\Ppsi        {\ensuremath{\psi}\xspace}                 
                  
 \mathchardef\PDelta="7101
 \mathchardef\PXi="7104
 \mathchardef\PLambda="7103
 \mathchardef\PSigma="7106
 \mathchardef\POmega="710A
 \mathchardef\PUpsilon="7107
                  
 \def\PB      {\ensuremath{B}\xspace}                 
                  
 \def\PD      {\ensuremath{D}\xspace}

 \def\PJ      {\ensuremath{J}\xspace}                 
 \def\PK      {\ensuremath{K}\xspace}

 \def\Pb      {\ensuremath{b}\xspace}                 
 \def\Pc      {\ensuremath{c}\xspace}

 \def\Pi      {\ensuremath{i}\xspace}

 \def\Ps      {\ensuremath{s}\xspace}

}

\def\mup        {\ensuremath{\Pmu^+}\xspace}
\def\mun        {\ensuremath{\Pmu^-}\xspace} 
\def\mumu       {\ensuremath{\Pmu^+\Pmu^-}\xspace}

\def\squark    {\ensuremath{\Ps}\xspace}

\def\cquark    {\ensuremath{\Pc}\xspace}

\def\bquark    {\ensuremath{\Pb}\xspace}

\def\pion  {\ensuremath{\Ppi}\xspace}
\def\piz   {\ensuremath{\pion^0}\xspace}

\def\pipi  {\ensuremath{\pion^+\pion^-}\xspace}

\def\kaon  {\ensuremath{\PK}\xspace}
  \def\Kbar  {\kern 0.2em\overline{\kern -0.2em \PK}{}\xspace}

\def\Kz    {\ensuremath{\kaon^0}\xspace}
\def\Kzb   {\ensuremath{\Kbar^0}\xspace}
\def\KzKzb {\ensuremath{\Kz \kern -0.16em \Kzb}\xspace}
\def\Kp    {\ensuremath{\kaon^+}\xspace}
\def\Km    {\ensuremath{\kaon^-}\xspace}

\def\KpKm  {\ensuremath{\Kp \kern -0.16em \Km}\xspace}
\def\KS    {\ensuremath{\kaon^0_{\rm\scriptscriptstyle S}}\xspace}

  \def\Dbar    {\kern 0.2em\overline{\kern -0.2em \PD}{}\xspace}
\def\D       {\ensuremath{\PD}\xspace}

\def\Dz      {\ensuremath{\D^0}\xspace}
\def\Dzb     {\ensuremath{\Dbar^0}\xspace}
\def\DzDzb   {\ensuremath{\Dz {\kern -0.16em \Dzb}}\xspace}
\def\Dp      {\ensuremath{\D^+}\xspace}
\def\Dm      {\ensuremath{\D^-}\xspace}

\def\DpDm    {\ensuremath{\Dp {\kern -0.16em \Dm}}\xspace}

\def\B       {\ensuremath{\PB}\xspace}
  \def\Bbar    {\kern 0.18em\overline{\kern -0.18em \PB}{}\xspace}

\def\Bu      {\ensuremath{\B^+}\xspace}

\def\Bd      {\ensuremath{\B^0}\xspace}
\def\Bs      {\ensuremath{\B^0_\squark}\xspace}
\def\Bsb     {\ensuremath{\Bbar^0_\squark}\xspace}
\def\Bdb     {\ensuremath{\Bbar^0}\xspace}

\def\jpsi     {\ensuremath{{\PJ\mskip -3mu/\mskip -2mu\Ppsi\mskip 2mu}}\xspace}

  \def\Y#1S{\ensuremath{\PUpsilon{(#1S)}}\xspace}

\def\L {\ensuremath{\PLambda}\xspace}
\def\Lbar {\ensuremath{\kern 0.1em\overline{\kern -0.1em\Lambda\kern -0.05em}\kern 0.05em{}}\xspace}

\def\BF         {{\ensuremath{\cal B}\xspace}}

\def\BR         {\BF}
\newcommand{\decay}[2]{\ensuremath{#1\!\to #2}\xspace}         

\def\to                 {\ensuremath{\rightarrow}\xspace}

\def\CP                {\ensuremath{C\!P}\xspace}

\def\BsToJPsiPhi  {\decay{\Bs}{\jpsi\phi}}

\def\AT#1     {\ensuremath{A_{\mathrm{T}}^{#1}}\xspace}           

\def\C#1      {\ensuremath{\mathcal{C}_{#1}}\xspace}                       
\def\Cp#1     {\ensuremath{\mathcal{C}_{#1}^{'}}\xspace}                    
\def\Ceff#1   {\ensuremath{\mathcal{C}_{#1}^{\mathrm{(eff)}}}\xspace}        
\def\Cpeff#1  {\ensuremath{\mathcal{C}_{#1}^{'\mathrm{(eff)}}}\xspace}       
\def\Ope#1    {\ensuremath{\mathcal{O}_{#1}}\xspace}                       
\def\Opep#1   {\ensuremath{\mathcal{O}_{#1}^{'}}\xspace}                    

\newcommand{\tev}{\ensuremath{\mathrm{\,Te\kern -0.1em V}}\xspace}
\newcommand{\gev}{\ensuremath{\mathrm{\,Ge\kern -0.1em V}}\xspace}
\newcommand{\mev}{\ensuremath{\mathrm{\,Me\kern -0.1em V}}\xspace}
\newcommand{\kev}{\ensuremath{\mathrm{\,ke\kern -0.1em V}}\xspace}
\newcommand{\ev}{\ensuremath{\mathrm{\,e\kern -0.1em V}}\xspace}
\newcommand{\gevc}{\ensuremath{{\mathrm{\,Ge\kern -0.1em V\!/}c}}\xspace}
\newcommand{\mevc}{\ensuremath{{\mathrm{\,Me\kern -0.1em V\!/}c}}\xspace}
\newcommand{\gevcc}{\ensuremath{{\mathrm{\,Ge\kern -0.1em V\!/}c^2}}\xspace}
\newcommand{\gevgevcccc}{\ensuremath{{\mathrm{\,Ge\kern -0.1em V^2\!/}c^4}}\xspace}
\newcommand{\mevcc}{\ensuremath{{\mathrm{\,Me\kern -0.1em V\!/}c^2}}\xspace}

\def\mum  {\ensuremath{\,\upmu\rm m}\xspace}

\def\invfb   {\ensuremath{\mbox{\,fb}^{-1}}\xspace}

\def\ps   {\ensuremath{{\rm \,ps}}\xspace}

\def\gsim{{~\raise.15em\hbox{$>$}\kern-.85em
          \lower.35em\hbox{$\sim$}~}\xspace}
\def\lsim{{~\raise.15em\hbox{$<$}\kern-.85em
          \lower.35em\hbox{$\sim$}~}\xspace}

\def\pt         {\mbox{$p_{\rm T}$}\xspace}

\def\evtgen     {\mbox{\textsc{EvtGen}}\xspace}
\def\pythia     {\mbox{\textsc{Pythia}}\xspace}

\def\geant      {\mbox{\textsc{Geant4}}\xspace}

\def\tell1  {TELL1\xspace}
\def\ukl1   {UKL1\xspace}

\usepackage{mciteplus}
\usepackage{styles}

\begin{document}
\renewcommand{\thefootnote}{\fnsymbol{footnote}}
\setcounter{footnote}{1}


\begin{titlepage}
\pagenumbering{roman}

\vspace*{-1.5cm}
\centerline{\large EUROPEAN ORGANIZATION FOR NUCLEAR RESEARCH (CERN)}
\vspace*{1.5cm}
\hspace*{-0.5cm}
\begin{tabular*}{\linewidth}{lc@{\extracolsep{\fill}}r}
\ifthenelse{\boolean{pdflatex}}
{\vspace*{-2.7cm}\mbox{\!\!\!\includegraphics[width=.14\textwidth]{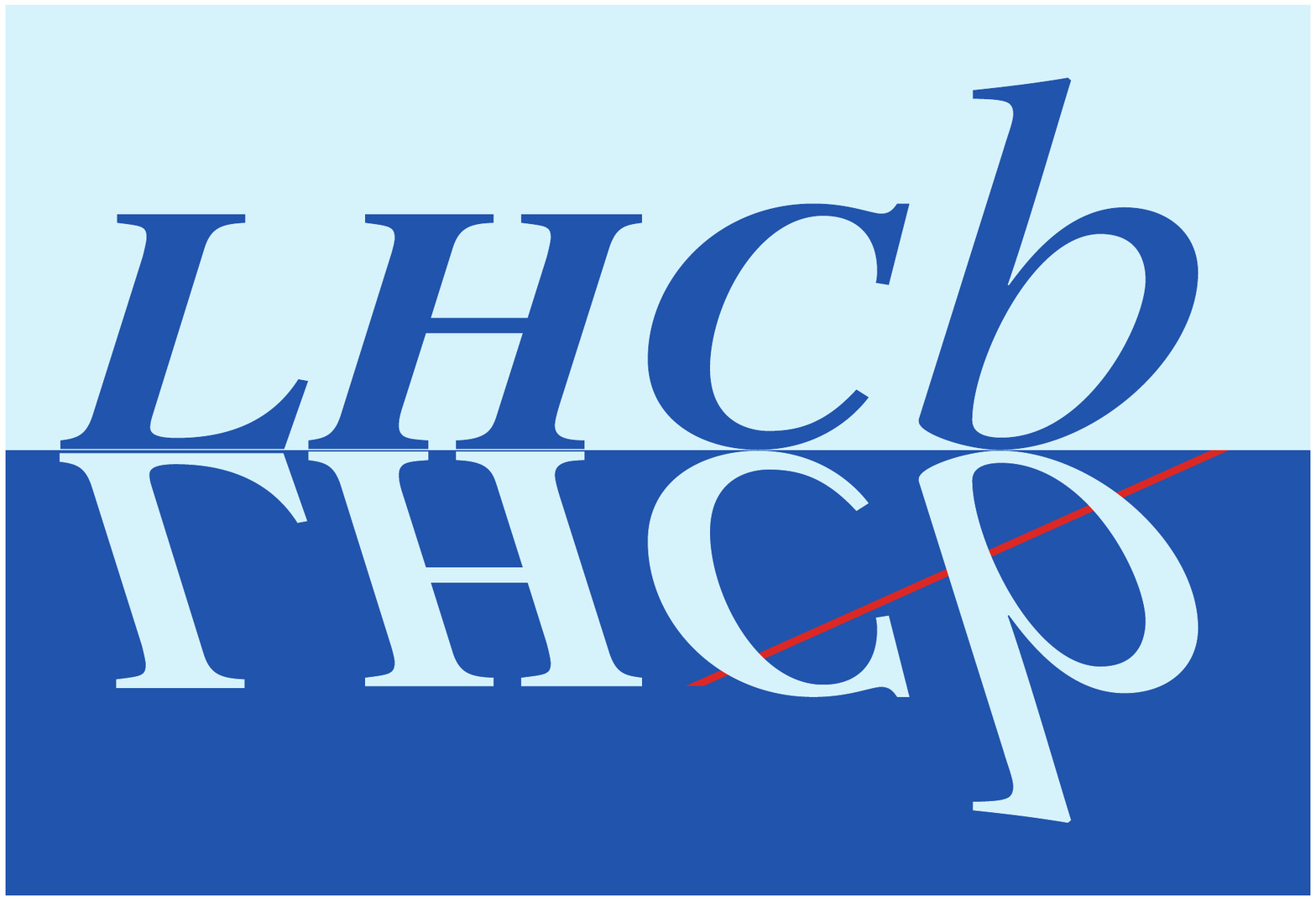}} & &}%
{\vspace*{-1.2cm}\mbox{\!\!\!\includegraphics[width=.12\textwidth]{figs/lhcb-logo.eps}} & &}%
\\
 & & CERN-PH-EP-2012-115 \\  
 & & LHCb-PAPER-2011-041 \\  
 & & 31 May 2012 \\ 
\end{tabular*}

\vspace*{4.0cm}

{\bf\boldmath\huge
\begin{center}
  Measurement of the \Bs\to\jpsi{}\KS branching fraction
\end{center}
}

\vspace*{2.0cm}

\begin{center}
The LHCb collaboration
\footnote{Authors are listed on the following pages.}
\end{center}

\vspace{\fill}

\begin{abstract}
  \noindent
The \Bs\to\jpsi{}\KS branching fraction is measured in a 
data sample corresponding to $0.41\:\invfb$ of integrated luminosity collected with the
LHCb detector at the LHC. This channel is sensitive to the penguin
contributions affecting the $\sin2\beta$ measurement from 
\Bd\to\jpsi{}\KS.
The time-integrated branching fraction is measured to be
$\BF(\Bs\to\jpsi\KS)=(\BsBF\pm\BsBFETT)\times10^{-5}$. 
This is the most precise measurement to date.
\end{abstract}

\centerline{(Submitted to Phys. Lett. B)}
\vspace*{2.0cm}
\vspace{\fill}

\end{titlepage}


\newpage
\setcounter{page}{2}
\mbox{~}
\newpage

\centerline{\large\bf LHCb collaboration}
\begin{flushleft}
\small
R.~Aaij$^{38}$, 
C.~Abellan~Beteta$^{33,n}$, 
B.~Adeva$^{34}$, 
M.~Adinolfi$^{43}$, 
C.~Adrover$^{6}$, 
A.~Affolder$^{49}$, 
Z.~Ajaltouni$^{5}$, 
J.~Albrecht$^{35}$, 
F.~Alessio$^{35}$, 
M.~Alexander$^{48}$
% ...

{\footnotesize \it
% ...
$ ^{5}$Clermont Universit\'{e}, Universit\'{e} Blaise Pascal, CNRS/IN2P3, LPC, Clermont-Ferrand, France\\
$ ^{6}$CPPM, Aix-Marseille Universit\'{e}, CNRS/IN2P3, Marseille, France\\
% ...
$ ^{33}$Universitat de Barcelona, Barcelona, Spain\\
$ ^{34}$Universidad de Santiago de Compostela, Santiago de Compostela, Spain\\
$ ^{35}$European Organization for Nuclear Research (CERN), Geneva, Switzerland\\
% ...
$ ^{38}$Nikhef National Institute for Subatomic Physics, Amsterdam, The Netherlands\\
% ...
$ ^{43}$H.H. Wills Physics Laboratory, University of Bristol, Bristol, United Kingdom\\
% ...
$ ^{48}$School of Physics and Astronomy, University of Glasgow, Glasgow, United Kingdom\\
$ ^{49}$Oliver Lodge Laboratory, University of Liverpool, Liverpool, United Kingdom\\
% ...
$ ^{n}$LIFAELS, La Salle, Universitat Ramon Llull, Barcelona, Spain\\
% ...
}
\end{flushleft}

\cleardoublepage


\renewcommand{\thefootnote}{\arabic{footnote}}
\setcounter{footnote}{0}



\pagestyle{plain} 
\setcounter{page}{1}
\pagenumbering{arabic}


%

\section{Introduction}\label{Sec:Introduction}
\begin{figure}[b]
\begin{minipage}[t]{0.49\textwidth}
\includegraphics[width=\textwidth]{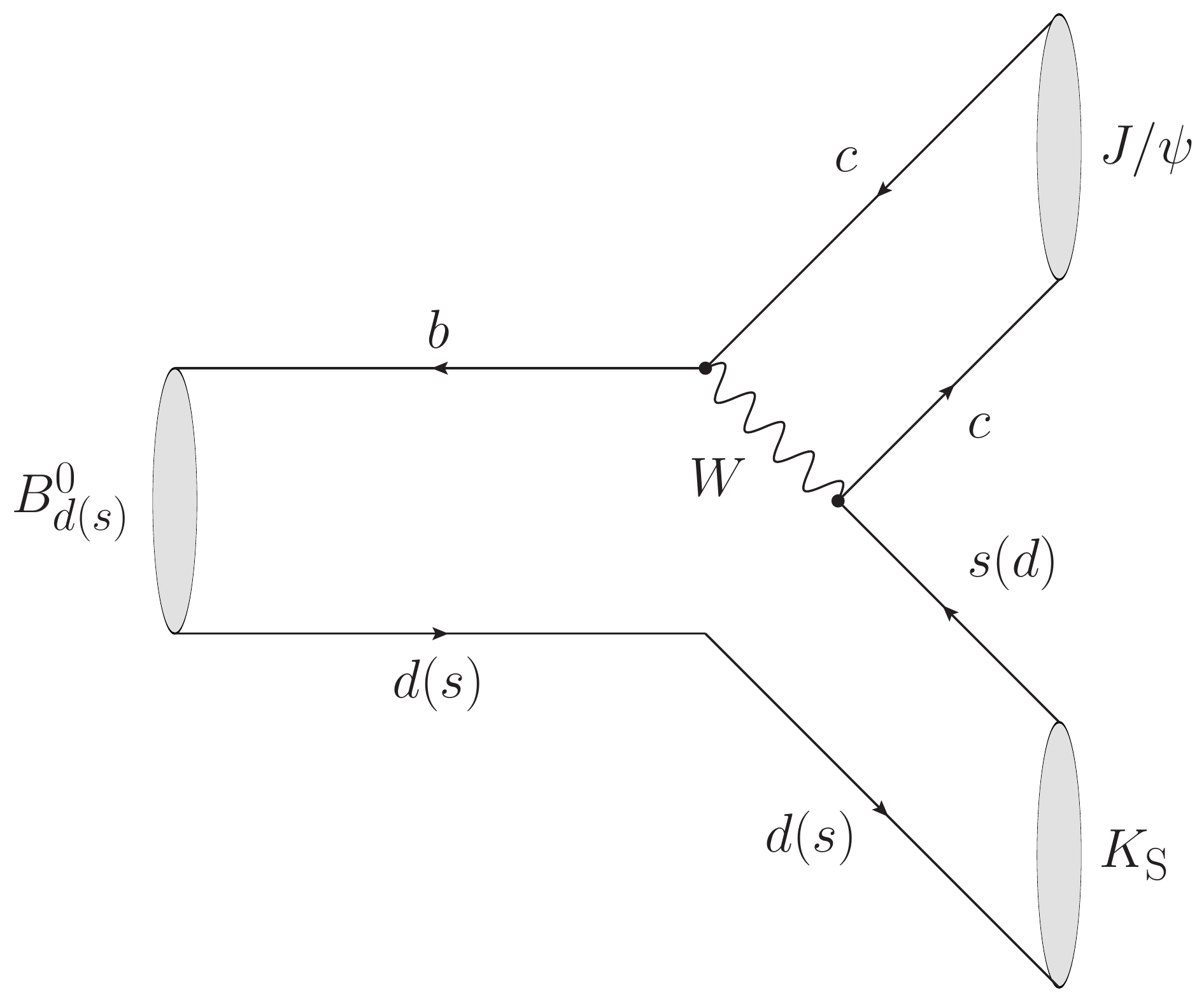}
\end{minipage}
\hskip 0.02\textwidth
\begin{minipage}[t]{0.49\textwidth}
\includegraphics[width=\textwidth]{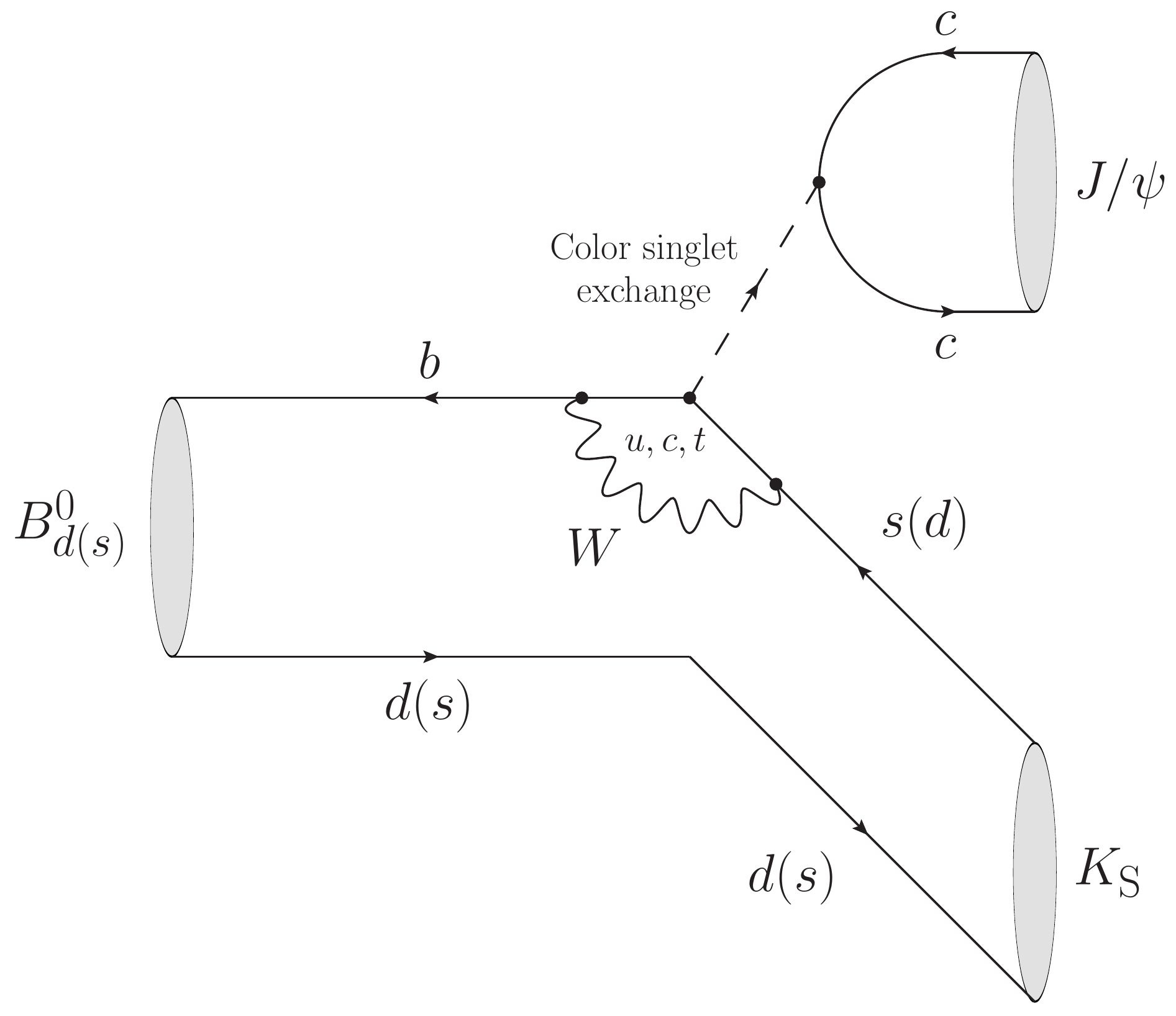}
\end{minipage}
\caption{Decay topologies contributing to the \Bd\to\jpsi{}\KS and \Bs\to\jpsi{}\KS channel: tree diagram to the left and penguin diagram to the right.\label{Fig:Feynman}}
\end{figure}

In the Standard Model (SM) \CP{} violation arises through a single
phase in the quark mixing matrix~\cite{Kobayashi:1973fv,*Cabibbo:1963yz}.  
In decays of neutral $\B$ mesons to a final state which is accessible to both $\B$ and $\Bbar$, 
the interference between the amplitude for the direct decay
and the amplitude for decay via oscillation leads to a
time-dependent \CP-violating asymmetry between the decay time
distributions of the two mesons. 
The mode \Bd\to\jpsi{}\KS allows for the measurement of such an asymmetry, which is 
parametrised by the $\Bd$--$\Bdb$ mixing phase $\phi_d$. In the SM this phase is equal to 
$2\beta$~\cite{Bigi:1981qs}, where $\beta$ is one of the angles of the unitarity triangle
of the mixing matrix.
This phase is already measured by the 
$B$ factories~\cite{Adachi:2012et,*Babar:2009yr} but an improved measurement 
is necessary to resolve conclusively the present tension in the 
unitarity triangle fits \cite{Faller:2008zc} and determine possible small deviations from the 
SM value.
To achieve the required precision, knowledge of the doubly Cabibbo-suppressed 
higher order perturbative corrections, known as \emph{penguin diagrams}, becomes mandatory.
The contributions of these penguin diagrams are difficult to calculate reliably and therefore need
 to be extracted directly from experimentally accessible observables.
Due to $SU(3)$ flavour symmetry, these penguin diagrams can be studied in other
decay modes where they are not suppressed relative to the tree level diagram.
The \Bs\to\jpsi{}\KS mode is the most promising candidate from the 
theoretical perspective since it is related to the \Bd\to\jpsi{}\KS mode
through the interchange of all $d$ and $s$ quarks 
($U$-spin symmetry, a subgroup of $SU(3)$)~\cite{Fleischer:1999nz}
and there is a one-to-one 
correspondence between all decay topologies in these modes, as illustrated in Fig.~\ref{Fig:Feynman}.
A further discussion regarding the theory of this decay and its 
potential at LHCb is given in Ref.~\cite{DeBruyn:2010hh,*DeBruyn:2010ge}.

To extract the parameters related to penguin contributions in these decays,
a time-dependent \CP violation study of the \Bs\to\jpsi{}\KS mode is required.
The measurement of its branching fraction is an important
first step, allowing to test the $U$-spin symmetry assumption that lies 
at the basis of the proposed approach.
The CDF collaboration reported the first observation of the \Bs\to\jpsi{}\KS 
decay~\cite{Aaltonen:2011sy}.
This letter presents a more precise measurement of this branching fraction at the LHCb experiment.

The strategy of the analysis is to measure the ratio of 
\Bs\to\jpsi{}\KS and \Bd\to\jpsi{}\KS event yields, which is
then converted into a \Bs\to\jpsi{}\KS branching fraction. We make
use of the \Bd\to\jpsi{}\Kz branching fraction and of the ratio of \Bs to \Bd meson
production at the LHC, denoted $f_s/f_d$~\cite{Aaij:2011hi,*Aaij:2011jp}.

We use an integrated luminosity of $0.41$\invfb{} of $pp$ collision
data recorded at a centre-of-mass energy of $7\tev$
during 2010 and the first half of 2011.  
The detector~\cite{Alves:2008zz} is a single-arm spectrometer designed
to study particles containing \bquark or \cquark quarks. 
It includes a high precision tracking system consisting of a
silicon-strip vertex detector surrounding the $pp$ interaction region,
a large-area silicon-strip detector located upstream of a dipole
magnet with a bending power of about $4{\rm\,Tm}$, and three stations
of silicon-strip detectors and straw drift-tubes placed
downstream. The combined tracking system has a momentum resolution
$\Delta p/p$ that varies from 0.4\% at 5\gevc to 0.6\% at 100\gevc,
and an impact parameter resolution of 20\mum for tracks with high
transverse momentum. Charged hadrons are identified using two
ring-imaging Cherenkov (RICH) detectors. Muons are identified by a muon
system composed of alternating layers of iron and multiwire
proportional chambers.

The signal simulation sample used for this analysis was generated using the
\pythia~\ensuremath{6.4} generator~\cite{Sjostrand:2006za} 
configured with the parameters detailed in Ref.~\cite{5873949}.
The \evtgen~\cite{Lange:2001uf}, \photos~\cite{Barberio:1993qi} 
and \geant~\cite{Agostinelli:2002hh} 
packages were used to decay unstable particles, 
generate QED radiative corrections and 
simulate interactions in the detector, respectively.

\section{Data samples and selection}\label{Sec:Data}
We search for \B\to\jpsi{}\KS decays\footnote{\B stands for \Bd or \Bs.}
where $\jpsi\to\mumu$ and
$\KS\to\pipi$.  Events are selected by a trigger system
consisting of a hardware trigger, which requires muon or hadron
candidates with high transverse momentum with respect to the beam
direction, \pt, followed by a two stage software 
trigger~\cite{LHCb-PUB-2011-016}. 
In the first stage a simplified event reconstruction is applied.  Events are
required to have either two oppositely charged muons with combined mass
above $2.7\:\gev/c^2$, or at least one muon or one high-\pt track ($\pt>1.8\:\gevc$) with a
large impact parameter with respect to any primary vertex.  In the second stage a
full event reconstruction is performed and only events 
containing \jpsi\to\mup{}\mun candidates are retained.

In order to reduce the data to a manageable level, very loose 
requirements are applied to suppress background 
while keeping the signal efficiency high.
\jpsi candidates are created from pairs of
oppositely charged muons that have a common vertex and a
mass in the range $3030$--$3150\:\mev/c^2$. The latter corresponds to
about eight times the $\mumu$ mass resolution at the \jpsi mass and covers
part of the $\jpsi$ radiative tail. 
The \KS selection requires two
oppositely charged particles reconstructed in the tracking stations 
on either side of the magnet, both with hits
in the vertex detector (long \KS candidate) or not (downstream \KS candidate).
The \KS candidates must be made of tracks forming a
common vertex and have a mass within eight standard deviations of the
$\KS$ mass and must not be compatible with
the \L mass under the mass hypothesis that one of the two tracks
is a proton and the other a pion.

We select \B candidates from combinations of \jpsi{} and \KS candidates
with mass $m_{\jpsi\KS}$ in the range $5200$--$5500~\mev/c^2$. The
latter is computed with the masses of the \mumu and \pipi pairs
constrained to the \jpsi and \KS masses, respectively.  
The mass and decay time of the
\B are obtained from a decay chain fit~\cite{Hulsbergen:2005pu} that in addition constrains the
\B candidate to originate from the primary
vertex.  The $\chi^2$ of the fit, which has
eight degrees of freedom, is required to be less than $128$ and
the estimated uncertainty on the \B mass must not exceed $30\:\mev/c^2$.
\B candidates are required to have a decay time larger than $0.2\:\ps$
and \KS candidates to have a flight distance larger than five times
its uncertainty. 
The offline selected signal candidate is required to 
be that used for the trigger decision at both software trigger stages.
About 1\% of the selected events have several 
candidates sharing some final state particles. In such cases one 
candidate per event is selected randomly.

\section{Measurement of event yields}\label{Sec:Analysis}
Following the selection described above, 
a neural network (NN) classifier~\cite{Feindt:2006pm} is used to further 
discriminate between signal and background. 
The NN is trained entirely on data, using samples that are independent of those
used to make the measurements. The training maximises the separation of 
signal and background events using weights determined 
by the \sPlot technique~\cite{Pivk:2004ty}. 
We use the \Bd\to\jpsi{}\KS signal in the data
as a proxy for the \Bs\to\jpsi{}\KS decay. The background events are
taken from  mass sidebands in the region 5390--5500\:$\mev/c^2$,
thus avoiding the \Bs signal region.
A normalisation  sample of one quarter of the candidates randomly selected
is left out in the NN training to allow an unbiased measurement of the \Bd yield.

\begin{figure}[t]\centering
\includegraphics[width=0.5\textwidth]{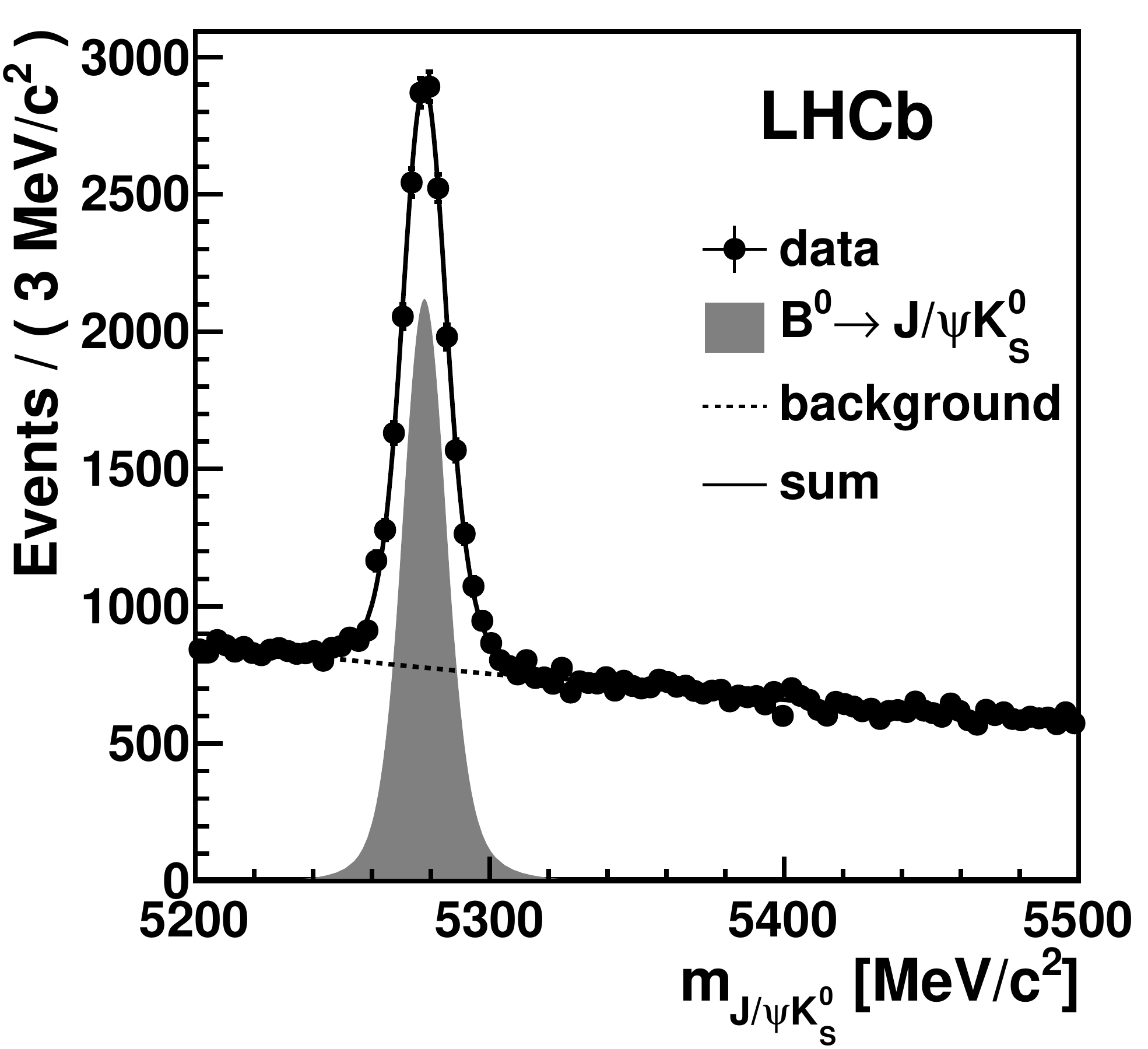}
\caption{Mass distribution of the \B\to\jpsi{}\KS candidates used to determine the PDF.
The solid line is the total PDF composed of the \Bd\to\jpsi{}\KS signal shown
in grey and the combinatorial background represented by the dotted line. \label{Fig:WeightingMassFit}}
\end{figure}
We perform an unbinned maximum likelihood fit to the mass distribution 
of the selected candidates, shown in Fig.~\ref{Fig:WeightingMassFit}, 
and use it to assign background and signal weights to each candidate.
The probability density function (PDF) is defined as the sum of a \Bd signal component, 
a combinatorial background
and a small contribution from partially reconstructed \B\to\jpsi{}\KS{}$X$ decays
at masses below the \Bd mass.
The mass lineshape of the \Bd\to\jpsi{}\KS signal in both data and  
simulation exhibits non-Gaussian tails on both sides of the signal peak
due to detector resolutions depending on angular distributions in the decay.
We model the signal shape by an empirical model composed of two 
Crystal Ball (CB) functions~\cite{Skwarnicki:1986xj}, one of which 
has the tail extending to high masses. 
The two CB components are constrained to have the same peak and width, 
which are allowed to vary in the fit. The parameters describing the CB tails are 
taken from \Bu\to\jpsi{}\Kp events which exhibit the same behaviour as \B\to\jpsi{}\KS.
The combinatorial background
is described by a second order polynomial. 
The \Bs\to\jpsi{}\KS signal is not included in this fit.
We extract $(14.4\pm0.2)\times10^3$ \Bd events from the fit.

The NN uses information about the candidate kinematics, vertex and track quality, 
impact parameter, particle identification information from the RICH and muon detectors, 
as well as global event properties like track and primary vertex multiplicities.
The variables that are used in the NN are chosen not to induce a correlation
with the  mass distribution. This was verified using simulated events.

To maximise the separation power, a first NN classifier using only the five most
discriminating variables is used to remove 80\% of the background events while
keeping 95\% of the \Bd signal. These variables
are the $\chi^2$ of the decay chain fit, the angle between the \B momentum and the 
vector from the primary vertex to the decay vertex, the \pt of the \KS, the estimated 
uncertainty on the \B mass and the impact parameter $\chi^2$ of the \jpsi.

The weighting procedure is then repeated on the remaining candidates and 
a second NN classifier containing 31
variables is trained. A cut is then made on the second NN output in order to optimise the 
expected sensitivity to the \Bs yield~\cite{Punzi:2003bu}.

\begin{figure}[tb]
\begin{minipage}[t]{0.49\textwidth}
\includegraphics[width=\textwidth]{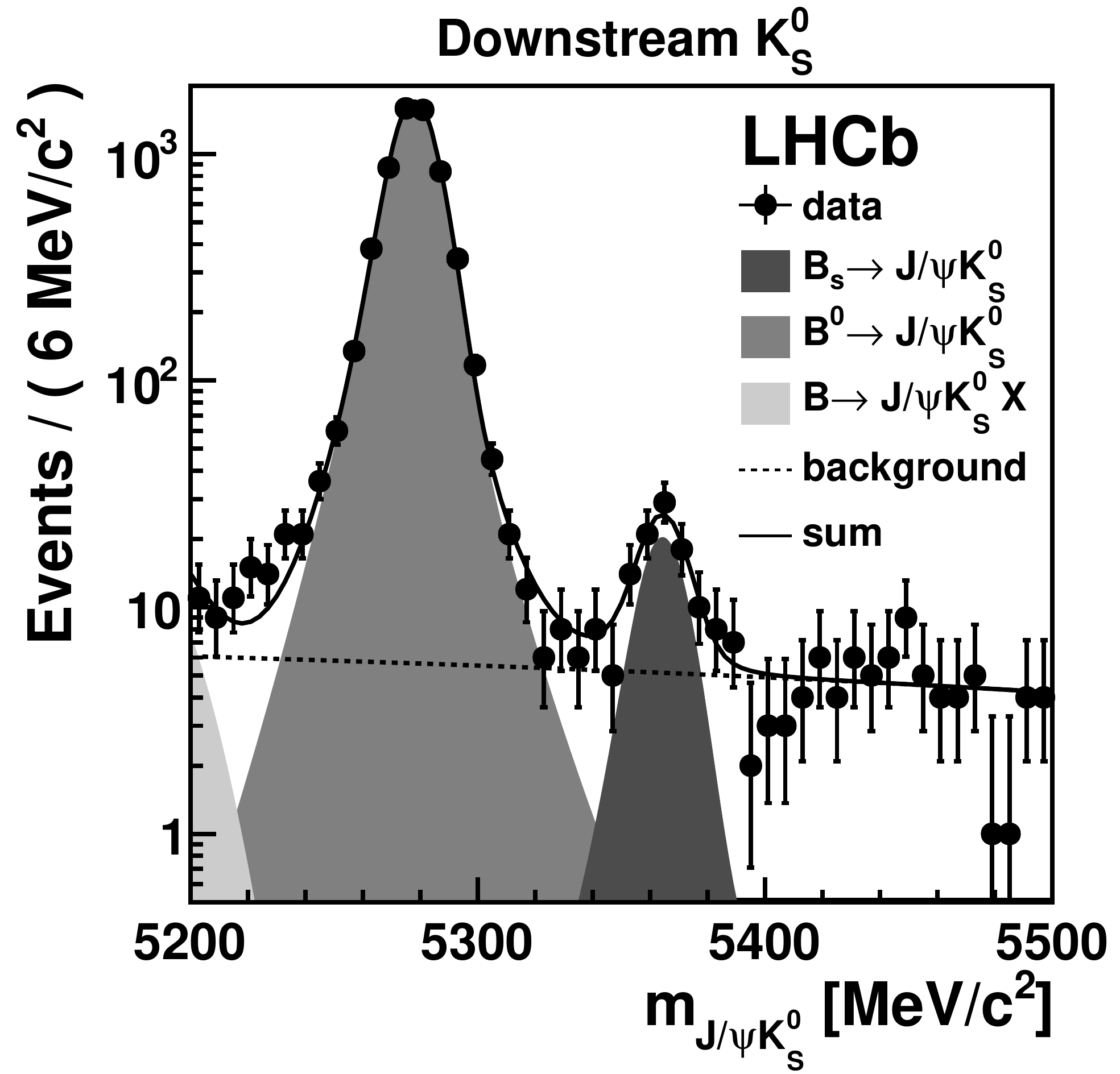}
\end{minipage}
\hskip 0.02\textwidth
\begin{minipage}[t]{0.49\textwidth}
\includegraphics[width=\textwidth]{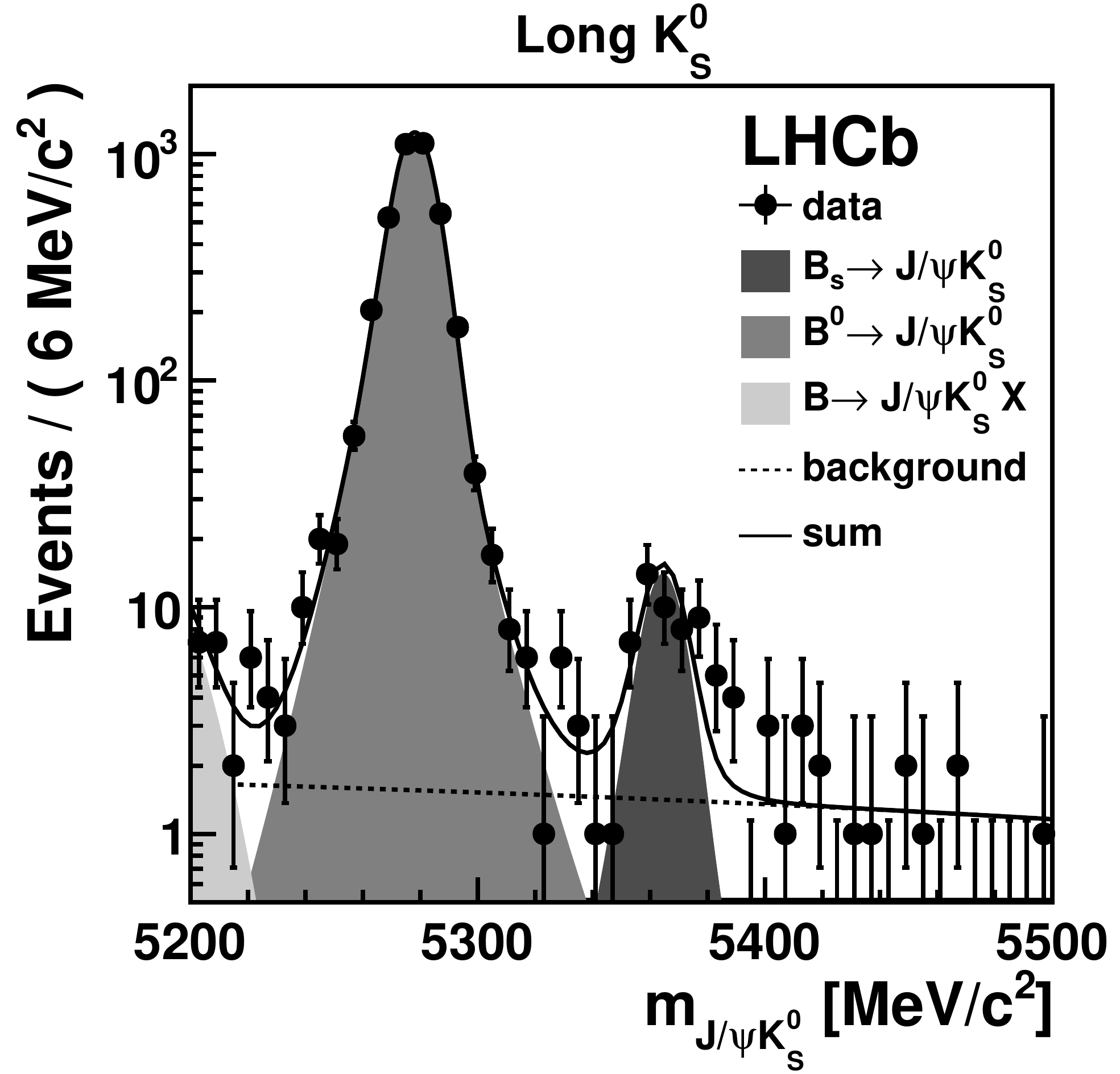}
\end{minipage}
\caption{Fit to full sample after the optimal NN cut has been applied with
  downstream \KS to the left and long \KS to the right.\label{Fig:BestFit}}
\end{figure}

For the candidates passing the NN requirement, we determine 
the ratio of \Bs and \Bd yields for candidates containing a downstream \KS or a long \KS
separately. The \Bd yield is measured in an unbinned likelihood fit 
to the normalisation sample and scaled to the full sample. 
The \Bs yield is fitted on the full sample. In both fits, the PDF is identical to that
used to determine the \sWeights with the addition of a PDF for the \Bs component, which
is constrained to have the same shape as the \Bd PDF, shifted by the measured
$\Bs-\Bd$ mass difference~\cite{Aaij:2011ep}. The results of the fits on the full samples
are shown in Fig.~\ref{Fig:BestFit} separately for candidates with downstream and long \KS.

The fitted yields are listed in Table~\ref{Tab:Yields}. The long and downstream
results are compatible with each other 
and are combined using a weighted average.

\begin{table}[!t]\begin{center}
\caption{\Bd and \Bs yields. Only statistical errors are quoted.
The \Bd yield is obtained in a fit to one quarter of the events
which have not been used in the NN training (normalisation sample)
and then scaled to the full sample.
\label{Tab:Yields}}
\vskip 1em
\begin{tabular}{l|P{5}|P{5}}
& \multicolumn{1}{c|}{ downstream \KS } & \multicolumn{1}{c}{ long \KS }\\
\hline
\Bd in normalisation sample                  & \DDBdNormFit!\DDBdNormFitE & \LLBdNormFit!\LLBdNormFitE \\
\Bd in normalisation sample (scaled to full) & \DDBdNorm!\DDBdNormE & \LLBdNorm!\LLBdNormE \\
\Bs in full sample                           & \DDBsFull!\DDBsFullE & \LLBsFull!\LLBsFullE\\
\hline
Ratio of \Bs to \Bd                          & \DDBsBd!\DDBsBdE & \LLBsBd!\LLBsBdE \\
\hline
\hline
Ratio of \Bs to \Bd (weighted average, $r$)       & \multicolumn{2}{C}{\BsBd\pm\BsBdE} \\
\end{tabular}
\end{center}\end{table}

\section{Corrections and systematic uncertainties}\label{Sec:Syst}
Differences in the total selection efficiencies between the \Bd\to\jpsi{}\KS 
and \Bs\to\jpsi{}\KS
arise because of the 
slight difference in momentum spectra of the \B mesons and/or the final state particles.
We find, using simulated events,
that the geometrical acceptance of the LHCb detector is lower
for the \Bs mode by $(1.3\pm0.5)\%$ where the error is due to the limited sample
of simulated events.
We correct for the ratio of acceptances and assign a 
conservative systematic uncertainty of 1.8\%, which
is the sum of the measured difference and its error.

The trigger, reconstruction and selection 
efficiencies also depend on the transverse momentum of the final
state particles. Applying the trigger transverse momentum cuts
on simulated \Bd and \Bs decays we find differences of up to 1\%,
 which is taken as systematic uncertainty.

Due to the selection cuts and the correlation of the neural network with the decay time,
a decay time acceptance function results in different selection efficiencies
for the \Bs and the \Bd. We determine the 
lifetime acceptance of the whole selection chain using simulated events, and find that the 
ratio of the time-integrated decay time distributions for \Bd and \Bs is $\BRdefSyst$.
The uncertainties on the parametrisation
of the lifetime acceptance cancel almost perfectly 
in the ratio, while the ones related to the \Bd and \Bs lifetimes and the
\Bs decay width difference $\Delta\Gamma_s$ do not.

The largest systematic uncertainty comes from the assumed mass PDF, 
in particular the fraction of the positive tail of the \Bd extending below the 
\Bs signal. We have studied the magnitude of this effect by leaving both tails of the 
CB shapes free in the fit, or by 
allowing the two CB shapes to have different widths. The maximal deviation we observe in the 
ratios of downstream or long candidates is 5\%, which we take as systematic uncertainty.
The effect of the uncertainty on the \Bs--\Bd mass difference is found to be $0.4\%$.

The corrections and systematic uncertainties affecting the branching fraction 
ratio are listed in Table~\ref{Tab:Syst}. The total uncertainty
is obtained by adding all the uncertainties in quadrature.

\begin{table}[tb]\begin{center}
\caption{Summary of corrections and systematic uncertainties on the ratio of branching fractions.\label{Tab:Syst}}
\begin{tabular}{l |C}
Source & \text{Correction factor} \\
\hline
Geometrical acceptance ($\epsilon_{\text{geom}}$) & \GeomSyst \\
Trigger and reconstruction  & \TrigSyst \\ 
Decay time acceptance  ($\epsilon_{\text{time}}$) &\BRdefSyst\\
Mass shape           & \FitSyst \\ 
\Bs-\Bd mass difference  & \BsBdSyst \\ 
\hline
Total & \eSyst\\
\end{tabular}
\end{center}\end{table}

We verify that the global event variable distributions, like the number
of primary vertices and the hit multiplicities,
are the same for \Bd and \Bs initial states
using the \BsToJPsiPhi channel.
We verify that the NN classifier is stable even when variables are removed from the training.
We search for peaking backgrounds in simulated  \bquark\to\jpsi{}$X$ 
events, and in data by inverting the \L veto and the \KS flight distance cut.
No evidence of peaking backgrounds is found.
All these tests give results 
compatible with the measured ratio though with a larger statistical uncertainty.

\section{Determination of branching fraction}\label{Sec:BR}
Using the measured ratio $r=\BsBd\pm\BsBdE$ of 
\Bs\to\jpsi{}\KS and \Bd\to\jpsi{}\KS yields, the 
geometrical ($\epsilon_{\text{geom}}$) and lifetime ($\epsilon_{\text{time}}$) acceptance ratios, and
assuming $f_s/f_d=0.267\aerr{0.021}{0.020}$~\cite{Aaij:2011hi,*Aaij:2011jp} we measure
the ratio of branching fractions
\begin{align}
  \frac{\BR(\Bs\to\jpsi\KS)}{\BR(\Bd\to\jpsi\KS)} & = r \times \epsilon_{\text{geom}} \times \epsilon_{\text{time}} \times
   \frac{f_d}{f_s}  \\
   & = 
  \BsBdR\pm\BsBdRE\text{\:(stat)}
  \pm\BsBdRS\text{\:(syst)}
  \pm\BsBdRfds{\:(f_s/f_d)}\nonumber
\end{align}
where the quoted uncertainties are statistical, systematic,  
and due to the  uncertainly in $f_s/f_d$, respectively.
Using the \Bd\to\jpsi{}\Kz branching fraction of $(8.71 \pm 0.32)\times 10^{-4}$~\cite{PDG},
we determine
the time-integrated \Bs\to\jpsi{}\KS branching fraction
\begin{align*}
  \BR(\Bs\to\jpsi\KS) = \left[\BsBF\right.&\pm \BsBFE\text{\:(stat)}
  \pm\BsBFEy\text{\:(syst)}
  \pm\BsBFEf{\:(f_s/f_d)} \\
  & \pm \left.\BsBFEp\text{\:(\BR(\Bd\to\jpsi{}\Kz))}
  \right]\times10^{-5}
\end{align*}
where the last uncertainty comes from the \Bd\to\jpsi{}\Kz branching fraction.
This result is compatible with, and more precise than, the previous 
measurement~\cite{Aaltonen:2011sy}.

\boldmath
\section{Comparison with $SU(3)$ expectations}\label{Sec:Interpretation}
\unboldmath
It was pointed out in Ref.~\cite{BRpaper} that because of 
the sizable decay width difference between the heavy and light eigenstates of the \Bs system,
there is an ambiguity in the definition of the branching
fractions of \Bs decays. Due to \Bs mixing, 
a branching fraction defined as the ratio of the time integrated 
number of \Bs decays to a 
final state and the total number of 
\Bs mesons, is not equal to the
\CP-average of the decay rates in the flavour eigenstate basis
\begin{equation}
  \BF(\Bs\to f)_{\text{theo}} = \frac{\tau_{\Bs}}{2}\left(\Gamma(\Bs\to f)+\Gamma(\Bsb\to f)\right)\big|_{t=0},
\end{equation}
used in the theoretical predictions; the restriction to $t=0$ removes the effects due to the 
non-zero $B_s$ decay width.
To obtain the latter quantity from the time-integrated decay rates the following correction factor
\begin{equation}\label{eq:BRDEF}
\frac{1-y_s^2}{1+{\cal A}^{\jpsi{}\KS}_{\Delta\Gamma}y_s} = \ExpToTheo,
\end{equation}
is applied,
where $y_s=\Delta\Gamma_s/2\Gamma_s$ is the normalised
decay width difference between the light and heavy states 
and ${\cal A}^{\jpsi{}\KS}_{\Delta\Gamma}$ is the final-state dependent 
asymmetry of the \Bs decay rates to the \jpsi{}\KS final state.
In calculating this correction factor we use 
$y_s  = \ysVal$~\cite{HFAG,*Asner:2010qj} and the SM expectation
$\mathcal{A}_{\Delta\Gamma_s}^{\jpsi{}\KS} = 0.84 \pm 0.18$~\cite{BRpaper}.

With this correction, and 
assuming $\BR(\Bs\to\jpsi\KS)_{\text{theo}}=\frac{1}{2}\BR(\Bs\to\jpsi\Kzb)_{\text{theo}}$
we get the \Bs\to\jpsi{}\Kzb branching fraction at $t=0$
\begin{align*} \label{eq:BFtheo}
  \BR(\Bs\to\jpsi\Kzb)_{\text{theo}} = (\TBsBF
  & \pm\TBsBFE\text{\:(stat)}
  \pm\TBsBFEy\text{\:(syst)} 
  \pm\TBsBFEf{\:(f_s/f_d)} \\ \nonumber
  & \pm\TBsBFEp\text{\:(\BR(\Bd\to\jpsi{}\Kz))}
  \pm\TBsBFEBR\:(y_s,\mathcal{A}_{\Delta\Gamma_s})
  )\cdot10^{-5}.
\end{align*}
This branching fraction can be compared to theoretical expectations from $SU(3)$ symmetry,
which implies an equality of the \Bs\to\jpsi{}\Kzb and \Bd\to\jpsi{}\piz 
decay widths~\cite{DeBruyn:2010hh}
\begin{equation}\label{Xi-def}
\Xi_{SU(3)}\equiv
\frac{\BR(\Bs\to\jpsi{}\Kzb)_{\text{theo}}}{2\BR(\Bd\to\jpsi{}\piz)}
\frac{\tau_{\Bd}}{\tau_{\Bs}}\frac{\left[m_{\Bd}\Phi(\Bd\to\jpsi{}\piz)\right]^3}
                               {\left[m_{\Bs}\Phi(\Bs\to\jpsi{}\Kzb)\right]^3}
\, \stackrel{SU(3)}{\longrightarrow}1,
\end{equation}
where the factor two is associated with the wave function of the \piz,
 $\tau_{B^0_{(s)}}$ is the mean $B^0_{(s)}$ lifetime and $\Phi$ refers to the 
two-body phase-space factors; see e.g. Ref.~\cite{Fleischer:1999nz}.

Taking the measured $\BR(\Bs\to\jpsi\Kzb)_{\text{theo}}$  and using the world 
average~\cite{PDG,Aaij:2011ep} for all other quantities, this ratio becomes
\begin{equation*}
\Xi_{SU(3)} = \XiSU \pm \XiSUE
\end{equation*}
and is consistent with theoretical expectation of unity under $SU(3)$ symmetry.

\section{Conclusion}\label{Sec:Conclusion}
The branching
fraction of the Cabibbo-suppressed decay \Bs\to\jpsi{}\KS is measured 
in a $0.41\:\invfb$ data sample collected with the LHCb detector. 
We determine the ratio
of the \Bs\to\jpsi{}\KS and \Bd\to\jpsi{}\KS branching fractions
to be $\frac{\BR(\Bs\to\jpsi\KS)}{\BR(\Bd\to\jpsi\KS)} = 
  \BsBdR\pm\BsBdRE\text{\:(stat)}
  \pm\BsBdRS\text{\:(syst)}
  \pm\BsBdRfds{\:(f_s/f_d)}.$
Using the world-average \Bd\to\jpsi{}\Kz branching fraction we 
get the time-integrated branching fraction
$\BF(\Bs\to\jpsi\KS)=[\BsBF\pm\BsBFE\text{\:(stat)}
  \pm\BsBFEy\text{\:(syst)}
  \pm\BsBFEf{\:(f_s/f_d)}
  \pm\BsBFEp\text{\:(\BR(\Bd\to\jpsi{}\Kz))}
  ]\times10^{-5}$. 
The total uncertainty of \BsBFETTR\ is dominated by the statistical uncertainty.
This branching fraction is compatible with expectations from $SU(3)$.

With larger data samples, a time dependent \CP-violation measurement of this
decay will be possible, allowing the experimental determination of the penguin contributions
to the $\sin2\beta$ measurement from \Bd\to\jpsi{}\KS.

\section*{Acknowledgements}

\noindent We express our gratitude to our colleagues in the CERN accelerator
departments for the excellent performance of the LHC. We thank the
technical and administrative staff at CERN and at the LHCb institutes,
and acknowledge support from the National Agencies: CAPES, CNPq,
FAPERJ and FINEP (Brazil); CERN; NSFC (China); CNRS/IN2P3 (France);
BMBF, DFG, HGF and MPG (Germany); SFI (Ireland); INFN (Italy); FOM and
NWO (The Netherlands); SCSR (Poland); ANCS (Romania); MinES of Russia and
Rosatom (Russia); MICINN, XuntaGal and GENCAT (Spain); SNSF and SER
(Switzerland); NAS Ukraine (Ukraine); STFC (United Kingdom); NSF
(USA). We also acknowledge the support received from the ERC under FP7
and the Region Auvergne.

\bibliographystyle{LHCb}
\bibliography{main}

\end{document}